\documentstyle[floats,epsf,aps]{revtex}
\epsfclipon

\begin{document}
\draft

\title{Fractal Conductance Fluctuations in Gold--Nanowires}

\twocolumn[\hsize\textwidth\columnwidth\hsize\csname @twocolumnfalse\endcsname

\author{Helmut Hegger$^1$, Bodo Huckestein$^2$, Klaus Hecker$^1$,
Martin Janssen$^2$, Axel Freimuth$^3$, Gernot Reckziegel$^1$,\\
R\"udiger Tuzinski$^4$}

\address{$^1$ II.\,Physikalisches Institut, Universit\"at zu K\"oln,
D-50937 K\"oln, Germany\\
$^2$ Institut f\"ur Theoretische Physik, Universit\"at zu K\"oln, 
D-50937 K\"oln, Germany\\
$^3$ Physikalisches Institut, Universit\"at Karlsruhe,
D-76128 Karlsruhe, Germany\\
$^4$ Institut f\"ur Halbleitertechnik der Technischen Hochschule
Aachen, D-52056 Aachen, Germany}

\date{\today{}}

\maketitle

\begin{abstract}

A detailed analysis of magnetoconductance fluctuations of
quasiballistic gold-nanowires of various lengths is presented.  We
find that the variance $\langle(\Delta\! G)^2\rangle = \langle
(G(B)-G(B+\Delta\! B))^2\rangle$ when analyzed for $\Delta\!  B$ much
smaller than the correlation field $B_c$ varies according to
$\langle(\Delta\! G)^2\rangle\propto \Delta\! B^{\gamma}$ with $\gamma
< 2$ indicating that the graph of $G$ vs.  $B$ is fractal.  We
attribute this behavior to the existence of long-lived states arising
from chaotic trajectories trapped close to regular classical orbits.
We find that $\gamma$ decreases with increasing length of the wires.

\end{abstract}


]

It is well established that quantum interference modifies the
conductance $G$ of disordered conductors and ballistic devices smaller
than the phase coherence length $L_\varphi $ (mesoscopic
regime)\cite{WW92}. In a semiclassical description one may view the
conductance electrons as moving along their classical trajectories,
i.e. ballistically between collisions. Since the phase information is
not lost over distances of order $L_\varphi $, a sample specific
interference pattern arises. This interference pattern may be altered
e.g. by applying a magnetic field $B$, which gives rise to
reproducible conductance fluctuations
(CF)~\cite{WW92,LS85,BILLIARDS1,BILLIARDS2,CHANG94,HH94,SCHEER94}.

It has recently been pointed out by Ketzmerick \cite{Ketz} that the
dwell time probability $P(t)$ for an electron to stay in the
mesoscopic sample longer than a time $t$ is related intimately to the
statistical properties of the CF. It is often assumed that the typical
time to cross the system, $t_D$, determines the behavior of $P(t)$ for
times larger than $t_D$, \begin{equation}
	P(t)\propto e^{-t/t_D} \label{1}\, .
\end{equation}
This form gives rise to a Lorentzian shape of the energy-dependent
conductance autocorrelation function $C(\Delta\! E) = \langle \delta
G(E) \delta G(E+\Delta\! E) \rangle$ with $\delta G(E)=G(E)-\langle
G(E)\rangle$ and correlation energy
$E_c=\hbar/t_D$\cite{BILLIARDS1,JBS90,BERRY94,B94}. Since $C(\Delta\!
E)= C(0)-0.5\langle(\Delta\! G)^2\rangle$ the variance
$\langle(\Delta\!  G)^2\rangle= \langle (G(E)-G(E+\Delta\!
E))^2\rangle$ behaves as $\langle(\Delta\! G)^2\rangle \propto
\Delta\! E^2$ for $\Delta\! E\ll E_c$.

The validity of Eq.~(\ref{1}) has been questioned by several authors
\cite{Ketz,AKL,Muz,Mirlin} and long-lived states have been predicted. If
$P(t)$ does not vary exponentially but decays algebraically,
\begin{equation}
	P(t)\propto t^{-\gamma} \; , \;\; 0<\gamma \leq 2 \label{2}
\end{equation} 
a non-trivial behavior of the variance appears\cite{Ketz}: 
\begin{equation}
\label{3} 
\langle(\Delta\! G)^2\rangle\propto \Delta\! E^{\gamma} \;\; {\rm for} \;\;\;
	\Delta\! E \ll E_c\, . 
\end{equation} 
It is important to note that exponents $\gamma $ in Eq.~(\ref{2})
which are larger than two cannot be detected in the variance. For
$\gamma >2$ the variance stays quadratic with respect to $\Delta \! E$
since analytic behavior dominates the lowest order approximation.

Measuring the length of the graph $G$ vs. $E$ on a scale $\Delta\!  E$
leads, as a consequence of Eq.~(\ref{3}), to a divergence proportional
to $(\Delta\! E)^{-(1-\gamma /2)}$, i.e.  the graph is fractal with
fractal dimension $D_F=2-\gamma/2$ \cite{Mandel}.

The conclusions discussed here for time $t$ and energy $E$ hold
similarly for other pairs of canonically conjugate variables, e.g. for
``area'' $A$ and magnetic field $B$, since for a closed path $A$ can
be considered as the accumulated area of an electron moving along the
path in time $t$.

In this letter we present measurements of the magnetoconductance
fluctuations of weakly disordered quasiballistic gold-nanowires with
various lengths $L<L_\varphi$. We adopt the method suggested by
Ketzmerick\cite{Ketz}. The analysis of $G(B)$ yields that $\langle
(\Delta \!  G^2) \rangle \propto \Delta \! B^\gamma $, where $\gamma $
is significantly smaller than two and decreases with increasing length
of the nanowire. We attribute this behavior to the existence of
long-lived states in the mesoscopic wire with a dwell time probability
$P(t)$ decaying much slower than exponentially. In addition we find
that the graph $G$ vs. $B$ is indeed fractal with fractal dimension
$D_F=2-\gamma/2$.
 
Gold-nanowires of very high purity (99.999\% ) were fabricated using
electron-beam lithography and lift-off with a four layer
polymethylmethacrylate- (PMMA) based resist system as described in
Ref.~\cite{LBMBW91}. The wires had a cross-section of 30 $\times $ 30
nm$^2$ and lengths between 400 nm and 1000 nm with a resolution of
$\pm 5 $ nm.  Current- and voltage-probes were connected to the wires
as shown in the inset of Fig.~4, providing a mesoscopic 2-probe
arrangement\cite{HH94}. The conductance $G(B)$ was measured as a
function of the magnetic field $B$ ($|B|\le 6$ T, 1 mT step-size) with
a standard ac lock-in technique in a $^3$He--$^4$He-dilution
refrigerator at $T \approx 60 $ mK\cite{HH94}.

The total resistance (nanowire and contacts) at $T = 60$ mK varies
between $R=8.2\; \Omega $ and $R= 18.3\; \Omega $ for $L = 400$ nm and
$L=1000$ nm, respectively. The resistance $R_c$ of the two
funnel-shaped contacts is $2.4\; \Omega $, as calculated from the
resistance per square $R_\Box = 0.5\; \Omega $.  From the residual
resistance ratio $R_{77\rm K}/R_{4.2\rm K}$ and from the electron
phonon inelastic scattering rate we obtain an elastic mean free path
$\ell $ in all investigated devices of $\ell \approx 60$ nm.  This is
larger than the width $W \simeq 30 $ nm and the thickness $T=30$ nm of
the wires and much larger than the Fermi-wavelength in gold
($\lambda_F \simeq 0.52 $ nm), i.e. our wires are in the
quasiballistic limit.  The phase-coherence length as determined from
weak localization at $T=100$ mK is $L_\varphi \approx 1.3\;\mu $m; the
spin-orbit scattering length is $L_{s.o.} \approx 0.2\; \mu
$m\cite{U91}.

\begin{figure}
\label{fig1}
  \begin{center}
    \leavevmode
    \epsfxsize=6.3cm
    \par
    \epsffile[18 77 545 730]{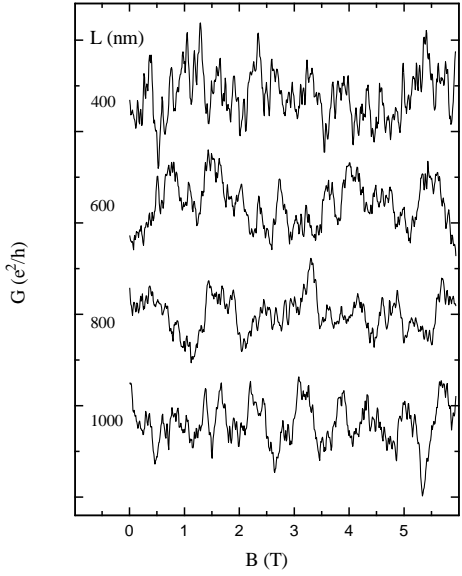}
    \par	
  \end{center}
\caption{Conductance $G(B)$ as a function of the magnetic field $B$
for 4 Au-wires with lengths as given in the figure. The traces are
shifted along the y-axis for clarity. (See text.)}
\end{figure}

In Fig.~1 we show $G(B) = G^{exp}(B) - G^{cl}(B)$ for four samples of
different lengths. Here $G^{exp}(B)$ is the conductance as measured at
60 mK. $G^{cl} \propto B^2$ is the weak classical magnetoconductance
obtained from a quadratic fit to $G^{exp}(B)$. The rms-amplitude of
the CF is of order ${\rm rms}(G) \approx 0.11\; e^2/h$ for all wires.

The correlation field $B_c$ as determined from the half-width of the
autocorrelation function $C(\Delta \! B) = \langle G(B) G(B+\Delta \!
B )\rangle$ ranges between 37 mT and 85 mT (see Tab.~1). The weak
dependence of $B_c$ on $L$ is related to the non-locality of the CF
and was discussed elsewhere~\cite{HH94}.

Before we analyze $\langle(\Delta \! G)^2\rangle$ we estimate the
contribution of the experimental noise to the measured conductance as
compared to that of the reproducible CF. We write the total
conductance as $G(B) = G_{cf}(B) + G_n(B)$, where $G_{cf}$ denotes the
reproducible part and $G_n$ is due to noise. Since $G(B)$ is measured
in a mesoscopic 2-probe configuration the CF are symmetric with
respect to reversal of the magnetic field,
i.e. $G_{cf}(B)=G_{cf}(-B)$.  On the other hand, the noise component
of a particular magnetoconductance trace may be viewed as being
composed of symmetric and antisymmetric parts $G_n^s$ and $G_n^a$,
where $G_n^s \equiv [ G_n(B) + G_n(-B) ]/2$, and $ G_n^a \equiv [ G_n(B) -
G_n(-B)]/2,$ so that $G_n^a(B)+G_n^s(B)=G_n(B)$, $G_n^s(B) =
G_n^s(-B)$, and $G_n^a(B) = - G_n^a(-B)$.  The antisymmetric part of the
total conductance $G^a = [G(B) - G(-B)]/2$ is a measure of the
antisymmetric noise component $G^a=G_n^a$. The symmetric part $G^s =
[G(B) + G(-B)]/2$ is given by $G^s=G_{cf}+ G_n^s$. Since it is
reasonable to assume that the symmetric and antisymmetric experimental
noise components are of equal magnitude the noise contribution to
$G^s$ (and thus to $G$) can be estimated from $G^a$.

\begin{figure}
\label{fig2}
  \begin{center}
    \leavevmode
    \epsfxsize=7.9cm
    \epsffile[0 0 700 521]{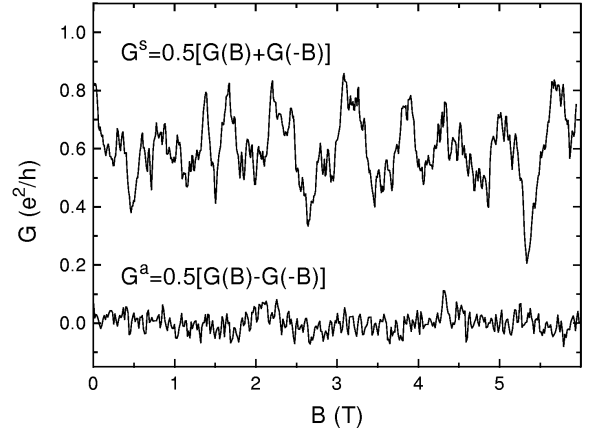}
  \end{center}
\caption{Symmetric $G^s$ and antisymmetric part $G^a$ of $G(B)$. $G^a$
is a measure of noise, whereas $G^s$ measures the reproducible
conductance fluctuations (see text). $G^s$ is shifted by $0.6e^2/h$
for clarity.}
\end{figure}

As an example we show in Fig.~2 $G^a$ and $G^s$ for the 1000 nm long
wire.  Similar results are obtained for all other wires. The
rms-noise-amplitude ${\rm rms}(G_n^a)\approx {\rm rms}(G_n^s)$,
calculated in the magnetic field range up to 6 T, is of order ${\rm
rms}(G_n^{s,a}) \leq 0.03\;e^2/h$ which is clearly smaller than the
amplitude of the reproducible CF amplitude ${\rm rms}(G)\approx 0.11\;
e^2/h$ (see Tab.~1). It should be noted that the rms-noise-amplitudes
${\rm rms}(G_n^a)$ calculated on magnetic field scales $\Delta \! B <
B_c$, which are relevant for the determination of $\gamma$, are much
smaller ($\leq 0.007\; e^2/h$). This is a reflection of the
1/f-component of the total experimental noise as function of the
magnetic field\cite{1_f}. The value for the rms-noise-amplitude
obtained from $G^a$ is in accordance with a more direct method where
we measured the system noise at a fixed magnetic field ($B=0$T).

We turn now to the calculation of $\langle (\Delta \! G)^2 \rangle $
as a function of $\Delta \! B $. We have used both, $G(B)$ and
$G^s(B)$ for this calculation. Figure 3 shows on a double logarithmic
scale $\langle (\Delta \! G)^2 \rangle $ as a function of $\Delta \!
B$. We also analyzed the experimental noise $\langle(\Delta \!
G^a)^2\rangle$ (inset of Fig.~3).  As shown, the values of the
derivative $d\ln\langle(\Delta \!  G)^2\rangle/d\ln(\Delta\! B)$ for
the noise $G^a$ are higher than those for $G^s$. A region,
approximately flat, which corresponds to a power law behavior with
$\gamma < 2$ can only be observed for $G^s$. Therefore, it is clear
that the oberservation of $\gamma < 2$ for $G^s$ is not caused by
experimental noise.

\begin{figure} 
\label{fig3} 
  \begin{center}
    \leavevmode
    \epsfxsize=7.5cm
    \epsffile[10 35 508 529]{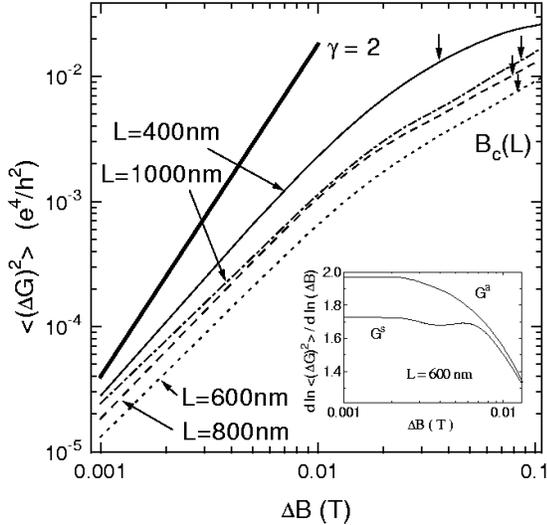}
  \end{center} 
\caption{Variance $\langle (\Delta \! G)^2 \rangle $
calculated from $G^s(B)$ as a function of $\Delta \!  B$ for the
gold-wires on a double logarithmic scale.  The correlation fields
$B_c$ are indicated by the arrows. The slope $\gamma $ of
$\langle(\Delta \!  G)^2\rangle $ versus $\Delta \! B$ has been
extracted by linear fits in the region $2-3\; {\rm mT} \leq \Delta \!
B \leq 10\; {\rm mT} \ll B_c$. For comparison we shows a thick solid
line of slope 2.  Inset: logarithmic derivative $d\ln\langle(\Delta \!
G^{s,a})^2\rangle/d\ln(\Delta\! B)$ for the nanowire with $L=600$ nm.}
\end{figure} 

We find that $\langle(\Delta\! G)^2\rangle$ varies as $\Delta
B^\gamma$ for $\Delta\! B \ll B_c$ with an uncertainty of the exponent
of $\pm 0.05$ (see Tab.~1).  Thus we conclude that at least for the
wires of length $L \geq 600$ nm $\gamma$ is definitely smaller than 2
which indicates long-lived states with a dwell time probability
$P(t)\propto t^{-\gamma}$. In addition we find that $\gamma$ decreases
with increasing system length $L$ of the nanowires (see Fig.~4) while
the signal to noise ratio increases.  We also performed a direct
fractal analysis on the graph $G$ vs. $B$ \cite{Mandel}. The results
obtained are in very good agreement with those analyzing the variance
and are listed in table 1. Measurements at higher temperatures ($T \le
4.2$ K) or on longer wires with ratio $L/L_\varphi \lesssim 2$ show
reduced CF and no significant change in the exponent $\gamma$.
 
For comparison we have also measured various disordered aluminum
nanowires with $\lambda_F\approx \ell \ll W,T$ and phase coherence
length $L_\varphi\approx 350\; {\rm nm} < L=500, 1000, 1500\; {\rm
nm}$. All the wires show values of $\gamma = 2$ within the
experimental uncertainty.

\begin{figure} 
\label{fig4} 
  \begin{center}
    \leavevmode
    \epsfxsize=6.8cm
    \epsffile[45 28 612 589]{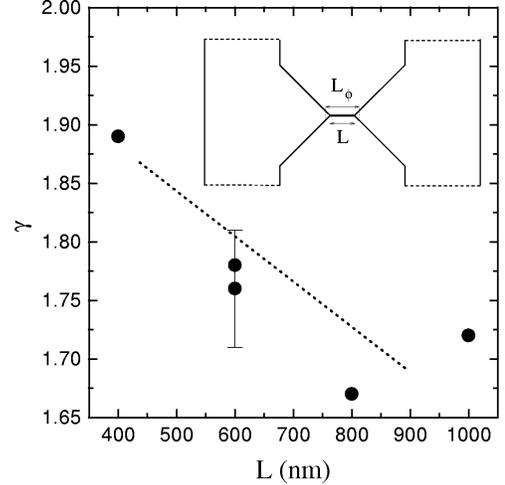}
  \end{center}
\caption{$\gamma $ versus $L$ (see text).  The inset shows the sample
layout, i.e. a Au-nanowire of length $L=1\mu$m with two funnel-shaped
current-voltage probes with an opening angle of 90$^\circ $.  This
corresponds to a mesoscopic 2-probe arrangement; current- and
voltage-leads are attached far outside the phase-coherence volume.
The dotted line serves as guide to the eye.}
\end{figure}

Now we turn to the discussion of our results.  As outlined in the
introduction the observation of $\gamma <2$ indicates the existence of
long-lived states. We know of two possible explanations for the
existence of long-lived states in mesoscopic conductors.  To
distinguish between them it is helpful to consider the classification
of mesoscopic conductors as introduced recently by Aleiner and Larkin
\cite{AL96}. A mesoscopic conductor for which the electronic path
between scattering events can be considered as purely classical was
denoted as being in the quantum chaos (QC) regime.  In this case, not
only the Fermi wavelength $\lambda_F$ has to be much smaller than the
(transport) mean free path $\ell$ but also the typical scale $d$ over
which the potential energy varies has to satisfy the constraint
$d^2>\lambda_F \ell$. Typical examples for such systems are antidot
arrays \cite{W93} where $d$ plays the role of the diameter of an
antidot and ballistic cavities where $d$ coincides with the size of a
cavity. Conductors with $d^2 < \lambda_F \ell$ are denoted as being in
the quantum disorder (QD) regime since the uncertainty $\delta x$ of
the electron position in the direction of its momentum after a
collision is larger than $d$.

It is now well known \cite{Ketz} that in the QC regime the
corresponding classical phase space is neither fully chaotic nor fully
integrable but contains in general islands of regular trajectories
with a self-similar structure.  Those classical chaotic trajectories
that spend long time close to regular orbits give rise to power law
behavior in the classical dwell time probability $P(t)\propto
t^{-\gamma}$. So far the prediction of the exponent $\gamma$ for a
given system does not seem to be possible. However, according to
Ketzmerick \cite{Ketz} several values for $\gamma$ which are of ${\cal
O }(1)$ have been reported for ballistic cavities as well as for
antidot arrays.

Another approach to the existence of long-lived states goes back to
the work of Altshuler et.al.(AKL) \cite{AKL}. They claim that
quasi-localized states already exist in the metallic regime of
mesoscopic conductors with a non-exponentially small
probability. These states are trapped by Bragg reflection in an
optimal potential fluctuation and should appear in the QC as well as
in the QD regimes.  They have attracted considerable interest in the
last two years and the results of AKL have been confirmed and improved
\cite{Muz,Mirlin}. The crucial result is that due to such
quasi-localized states the dwell time probability (in a wire) should
behave for not to large times $t\gg t_D$ as: \begin{equation} P(t)
\propto \exp\left( -g \ln^2 (t/t_D)\right) \propto t^{-g\ln (t/t_D)}\,
.  \end{equation} Here $g=Gh/e^2$ is the dimensionless conductance
which is large in a metallic wire, $g \gg 1$ (see Tab.~1). Thus one
cannot detect the corresponding long-lived states in the CF which are
only sensitive to a power law behavior in $P(t)$ with an exponent
$\gamma$ smaller than two.

\begin{table}[tb]
\begin{tabular}{ccccccc}
\label{tab1}
\narrowtext
 $L$ \hspace{0.1cm}
 & $G$ \hspace{0.1cm}
 & ${\rm rms}(G)$ \hspace{0.1cm}
 & $B_c$ \hspace{0.1cm}
 & \hspace{0.2cm} $\gamma^\star $ \hspace{0.2cm}
 & $\gamma$ 
 & $\gamma_F$ \\
  ($\mu $m) \hspace{0.1cm}
 & $(e^2/h)$ \hspace{0.1cm}
 & $(e^2/h)$ \hspace{0.1cm}
 & (mT) \hspace{0.1cm}
 & \hspace{0.2cm} \hspace{0.2cm}
 & 
 & \\ 
\tableline 
0.4 & 3135 & 0.105 & 37 & 1.86 & 1.89 & 1.88\\ 
0.6$a$ & 1968 & 0.13 & 80 & 1.77 & 1.78 & 1.75\\ 
0.6$b$ & 1645 & 0.12 & 80 & 1.68 & 1.76 & 1.74\\ 
0.8 & 1286 & 0.1 & 78 & 1.75 & 1.68 & 1.68\\ 
1.0 & 1408 & 0.1 & 85 & 1.74 & 1.72 & 1.70\\ 
\end{tabular} 
\narrowtext
\caption{$L$: Length of the nanowires; $G$: Total conductance; ${\rm
rms}(G)$: Root-mean-square of the CF with an
accuracy of $\pm 0.01\; e^2/h$; $B_c$: Correlation field extracted
from the half-width of the autocorrelation function $C(\Delta\! B)$
with an accuracy of $\pm 10 $ mT; $\gamma^\star$ and $\gamma$ have
been extracted by fits of the variance $\langle (\Delta \! G)^2
\rangle $ calculated from $G(B)$ and $G^s(B)$, respectively with an
uncertainty of $\pm 0.05$. $\gamma_F$ was obtained by a direct fractal
analysis of the graph $G$ vs. $B$ with an uncertainty of $\pm 0.04$.}
\end{table}

Turning now to our experiments we note that our gold-wires are
characterized by an elastic mean free path which is caused mainly by
diffusive boundary scattering, $\ell > W,T$.  In addition there is
also a certain amount of specular boundary scattering and one may
classify the samples as QC conductors with $d\approx W$. We therefore
conclude that the observed behavior of the variance $\langle(\Delta\!
G)^2\rangle \propto \Delta \!  B^{\gamma}$ in the Au-wires indicates
the existence of long-lived states with a dwell time probability
$P(t)\propto t^{-\gamma}$, $\gamma < 2$.  In contrast, the aluminum
wires are definitely in the QD regime, since $\lambda_F\approx \ell$,
consistent with $\gamma \simeq 2$.

In summary, we have presented measurements of magnetoconductance
fluctuations in mesoscopic gold-wires, which are in the quantum chaos
regime. Analysis of the CF-pattern reveals power law behavior of the
variance $\langle(\Delta\! G)^2\rangle = \langle (G(B)-G(B+\Delta\!
B))^2\rangle \propto \Delta\! B^\gamma$ with $\gamma < 2$ for small
magnetic field intervals $\Delta\! B\ll B_c$. This non-analyticity of
$\langle(\Delta\! G)^2\rangle$ is a signature of long-lived states in
the samples. Such states are a generic feature of systems with a mixed
(chaotic and regular) phase space of the corresponding classical
system. Additional measurements, e.g. of the temperature dependence of
$\gamma$, should provide further insight into the origin of the
fractal conductance fluctuations.

We thank R.~Sch\"afer, B.~B\"uchner, R.~Ketzmerick, J.~ Hajdu and
H.~Micklitz for useful discussions. This work was supported by the
Deutsche Forschungsgemeinschaft through SFB 341.


\begin{references}

%\bibitem[*]{hh:address} email (H.~Hegger): hh@mesofix.ph2.uni-koeln.de

\bibitem{WW92} S. Washburn and R. A. Webb, Rep. Prog. Phys. {\bf 55},
1311 (1992), and references therein.

\bibitem{LS85} P. A. Lee and A. D. Stone, Phys. Rev. Lett. {\bf 55},
1662 (1985); B. L. Altshuler, JETP Lett. {\bf 41}, 648 (1985);
P. A. Lee, A. D. Stone, and H. Fukuyama, Phys. Rev. B {\bf 35}, 1039
(1987).

\bibitem{BILLIARDS1} C. M. Marcus, A. J. Rimberg, R. M. Westervelt,
P. F. Hopkins, and A. C. Gossard, Phys. Rev. Lett. {\bf 69}, 506 (1992).

\bibitem{BILLIARDS2} C. M. Marcus, R. M. Westervelt, P. F. Hopkins, and
A. C. Gossard, Phys. Rev. B {\bf 48}, 2460 (1993).

\bibitem{CHANG94} A. M. Chang, H. U. Baranger, L. N. Pfeifer, and
K. W. West, Phys. Rev. Lett. {\bf 73}, 2111 (1994).

\bibitem{HH94} K. Hecker, H. Hegger, R. Sch\"afer, U. Murek,
C. Braden, and W. Langheinrich, Phys. Rev. B {\bf 50}, 18601 (1994).

\bibitem{SCHEER94} E. Scheer, H. v. L\"ohneysen, and H. Hein,
Physica B {\bf 218}, 85 (1996).

\bibitem{Ketz} R. Ketzmerick, preprint cond-mat/9510007 (1995),
accepted for publication in Phys. Rev. B, Oct. 15 (1996), and
references therein.

\bibitem{JBS90} R. A. Jalabert, H. U. Baranger, and A. D. Stone,
Phys. Rev. Lett. {\bf 65}, 2442 (1990); H. U. Baranger,
D. P. DiVincenzo, Phys. Rev. {\bf B}, 10637 (1991); H. U. Baranger,
R. A. Jalabert, and A. D. Stone, Phys. Rev. Lett. {\bf 70}, 3876
(1993).

\bibitem{BERRY94} M. J. Berry, J. A. Katine, R. M. Westervelt, and
A. C. Gossard, Phys. Rev. B {\bf 50}, 17721 (1994).

\bibitem{B94} M. V. Budantsev, Z. D. Kvon, A. G. Pogosov,
L. V. Litvin, V. G. Mansurov, V. P. Migal, S. P. Moshchenko, and
Yu. Nastaushev, JETP Lett. {\bf 59}, 645 (1994).

\bibitem{AKL} B. L. Altshuler, V. E. Kravtsov, and I. V. Lerner, JETP
Lett. {\bf 45}, 199 (1987); in {\it Mesoscopic Phenomena in Solids},
edited by B. L. Altshuler, P. A. Lee, and R. A. Webb (North-Holland,
Amsterdam, 1991), p.~449.

\bibitem{Muz} B. A. Muzykantskii and D. E. Khmelnitskii, Phys. Rev. B
{\bf 51}, 5480 (1995); B. A. Muzykantskii and D. E. Khmelnitskii,
preprint cond-mat/9601045 (1996), (unpublished).

\bibitem{1_f} The fourier transform of $G^a(B)$ shows a 1/f-noise
component for frequencies corresponding to magnetic field changes
$\Delta B \geq 12$mT. 

\bibitem{Mandel} B. B. Mandelbrot, {\it The Fractal Geometry of
Nature} (Freeman, San Francisco, 1982).

\bibitem{LBMBW91} W. Langheinrich, H. Beneking, U. Murek, C. Braden,
and D. Wohlleben, J. Vac. Sci. Technol. B {\bf 9}, 2904 (1991).

\bibitem{U91} U. Murek, thesis, University of Cologne (1991),
(unpublished).

\bibitem{AL96} I. L. Aleiner and A. I. Larkin, preprint
cond-mat/9603121 (1996), (unpublished).

\bibitem{W93} D. Weiss, M. L. Roukes, A. Menschig, P. Grambow,
K. von Klitzing, and G. Weimann, Phys. Rev. Lett {\bf
66}, 2790 (1991).

\bibitem{Mirlin} A. D. Mirlin, JETP Lett. {\bf 62}, 603 (1995).

\end{references}
\end{document}